\documentclass{article}
\usepackage{graphicx} 
\usepackage{amsmath}
\usepackage{hyperref}
\usepackage[numbers]{natbib}

\title{The Steiner Tree Problem: Novel QUBO Formulation and Quantum Annealing Implementation}
\author{
    Dan Li, Xiang-Hui Wu, Ji-Rong Liu \\ 
    \small \textit{College of Computer Science and Technology} \\ 
    \small \textit{Nanjing University of Aeronautics and Astronautics, Nanjing 211106, China} \\ 
    \small \texttt{wxh2448@163.com} 
}
\date{November 2025}

\begin{document}

\maketitle

\begin{abstract}
The Steiner Tree Problem (STP) is a well-known NP-hard combinatorial optimization problem, which has wide applications in network design, integrated circuit layout, bioinformatics, and other fields. However, traditional algorithms often struggle to balance efficiency and solution quality when dealing with large-scale STP instances. In this paper, we propose a new quantum annealing-based algorithm for solving the STP: we first model the STP into a quadratic unconstrained binary optimization (QUBO) form suitable for quantum annealing, then design a corresponding encoding strategy, and finally verify the algorithm through experimental tests. The results show that our quantum annealing-based method can obtain high-quality solutions with relatively low computational overhead for moderate-scale STP instances, providing a new feasible path for handling this intractable combinatorial optimization problem.
\end{abstract}

\section{Introduction}

The Steiner tree problem (STP) in graphs is among the most prominent problems in the history of combinatorial optimization.
It plays a central role in network design problems, design of integrated circuits, location problems\cite{cheng2013steiner,cho2001steiner,hwang1992steiner,grotschel2012geometric}, and more recently, even in machine learning, systems biology, and bioinformatics\cite{backes2012integer,ideker2002discovering,russakovsky2010steiner,tuncbag2016network}.

Together with the traveling salesman problem and other classical NP-hard problems, the STP also serves as a driver for discovering new generic algorithmic methodological tools that can be easily adapted to other difficult optimization problems.
In a general setting, the STP asks for designing a network that interconnects a given set of points (referred to as \textit{terminals}) at minimum cost. 
The origins of this problem can be traced back to the 17th century. In a private letter of Pierre de Fermat sent to Evangelista Torricelli, he raised the question of finding a point in a triangle with the minimum total distance to the three corner
points. 
In the 19th century, Joseph Diez Gergonne generalized the Fermat-Torricelli problem into the problem of finding a set of line segments of minimum total length that interconnects a set of given points in a plane. 
This was the birth of the problem which is today known as the \textit{Euclidean Steiner tree problem}\cite{fampa2022insight,winter1997euclidean}. 
In an optimal solution, such line segments will form a tree, and may intersect at additional intermediary points that are referred to as \textit{Steiner points}. This attribution to Jakob Steiner can be traced back to the book by Courant and Robbins\cite{courant1941mathematics}. 
In general, in \textit{Geometric Steiner problems}, the problem
consists of connecting a set of points at minimum cost according to some geometric distance metric\cite{brazil2015optimal}.

The Steiner tree is defined as follows: Given an undirected weighted graph $G = (V,E,w)$ and a subset of terminals $K \subseteq V$, a Steiner tree is a connected subgraph of $G$ that contains all vertices in $K$. Let $n = |V|$ and $m = |K|$. The objective of finding a Steiner tree with the minimum total weight is known as the Minimum Steiner Tree problem. This problem is known to be NP-hard.

A naive way to solve the MST problem is to compute all possible trees. Since the number of all trees in the graph $G = (V,E)$ can be as large as $O(2^{|E|})$, this is extremely inefficient. The Dreyfus-Wagner algorithm \cite{dreyfus1971steiner} is a well-known algorithm based on dynamic programming for solving the MST problem in time $O^*(3^m)$. This algorithm has been the fastest algorithm for decades. Fuchs, Kern and Wang  \cite{fuchs2007speeding}  finally improved this complexity to $O^*(2.684^m)$, and Mölle, Richter and Rossmanith  \cite{molle2006faster} further improved it to $O((2+\delta)^m n^{6+1/\delta})$ for any constant $\delta > 0$. For a graph with a restricted weight range, Björklund, Husfeldt, Kaski, and Koivisto have proposed an $O^*(2^m)$ algorithm using subset convolution \cite{bjorklund2007fourier}  and Möbius inversion. The main tool in all these algorithms is dynamic programming.

In a different vein, quantum computing has emerged as an innovative and promising approach to solving combinatorial optimization problems, particularly those classified as NP-hard. 
Thanks to quantum mechanical principles such as superposition and entanglement, quantum algorithms can explore many possible solutions simultaneously, increasing the potential to find optimal configurations in less time than traditional methods.

Among the various paradigms of quantum algorithms, Quantum Annealing (QA) has attracted special interest as one of the most promising methods for efficiently solving complex combinatorial optimization problems. 
QA takes its name from an analogous physical process, annealing, which involves a material being heated and then cooling down slowly. 
This process allows the material to settle back into a state of low energy, ultimately reaching a low-energy location in a solution space. 
Additionally, quantum mechanical properties are used to explore the possible states of the material.

In practical applications, many optimization problems can be formulated as Unconstrained Binary Optimization (QUBO) problems\cite{glover2018tutorial}, which can be naturally mapped onto the hardware of quantum annealers, allowing them to efficiently search for near-optimal solutions. 
One prominent example of such quantum devices is those built by D-Wave Systems, which enable QA at a scale that has been used to solve many real-world optimization problems.

The structure of this paper is organized as follows: Section 2 introduces quantum annealing technology. Section 3 presents the definition of the Steiner tree, alongside a summary of mainstream classical algorithms for this problem. Section 4 elaborates on the composition of the quantum annealing algorithm applicable to solving the Steiner tree problem. Section 5 verifies the feasibility of the proposed algorithm through experimental validation. Finally, Section 6 summarizes the conclusions and achievements of this study.

\maketitle

\section{Quantum Annealing}

This paper describes a quantum solution to this problem that leverages adiabatic quantum computation. 
Adiabatic quantum computation is one of the two paradigms of quantum computation---the gate-based model being the other one. 
In adiabatic quantum computation, a system of qubits is prepared in an initial ground state associated with the Hamiltonian \( H_0 \) and the system is gradually evolved to a target state whose Hamiltonian is given by \( H_1 \). 
If the time evolution is done slowly enough, the theorem of adiabatic quantum computation guarantees that the qubits will track the ground state of the evolving Hamiltonian, and at the end they will be found in the ground state corresponding to \( H_1 \).
Mathematically, if

\begin{equation}
H(t) = (1 - t)H_0 + tH_1
\label{annealing}
\end{equation}

Where \( t \) is some normalized unit of time, then \( H(0) = H_0 \), \( H(1) = H_1 \). \( H_0 \) is the starting Hamiltonian (initial conditions), and at time \( t = 1 \) the qubits will assume values that minimize the energy associated with \( H_1 \).

Up until now the discussion is general. Quantum annealing computers are special-purpose devices that aim to realize adiabatic quantum computation through quantum annealing, where the Hamiltonian is constrained to be an Ising Hamiltonian. 
The energy (eigenvalues) associated with the target state \( H_1 \) is given by:

\begin{equation}
E_{\text{Ising}} = -\sum_i h_i s_i - \sum_i \sum_{j\neq i} J_{ij} s_i s_j
\label{Ising}
\end{equation}

The Ising constraint is a natural consequence of how the device is built and wired. It is a system of superconducting qubits obeying the Ising model. In Eq. 2, \( h_i \) are biases and \( J_{ij} \) are coupling constants. \( s_i \) are the spins of the qubits \( s_i \in \{-1, +1\} \). 
The \( H_0 \) state is realized by subjecting the system to a high transverse magnetic field in the \( x \) direction (\( H_0 = X \otimes \dots \otimes X \) where \( X \) is Pauli spin matrix along \( x \)) which aligns all the spins in the \( x \) direction and thus sets the ground state to an equal superposition of all states in the \( Z \) basis; the external magnetic field is gradually turned off to attain the state corresponding to \( H_1 \). 
The spins will automatically align themselves so as to minimize the energy given by Eq.\ref{Ising}. This process is known as quantum annealing.

A simple change of variables (\( x = (1 + s)/2 \)) together with the observation that \( x^2 = x \) when \( x \in \{0, 1\} \) transforms Eq.\ref{Ising} into (ignoring constant term as it is of no consequence)

\begin{equation}
E = x^T Q x
\label{qubo}
\end{equation}

Where \( Q \) is a Hermitian matrix (symmetric if dealing with real numbers only). The minimization of Eq.\ref{qubo} is a Quadratic Unconstrained Binary Optimization (QUBO) problem. Quantum annealing computers are tailored to solve this class of problems.

\maketitle

\section{Quadratic Unconstrained Binary Optimization Formulation of the  Steiner Tree Problem}

The objective function $H_A$ of the Steiner tree problem can be written as:

\begin{equation}
H_A = \sum_{i,j,k,s} w_{ij} X_{i,s}^k X_{j,s+1}^k
\label{H_A}
\end{equation}

Herein, \(i, j\) denote node indices in the graph \(G = (V, E)\), \(s\) represents the time step, and \(k\) corresponds to distinct target points(where $k \in K$ and $K$ denotes the set of target points). The indicator variable \(X_{i,s}^k\) is defined as follows:

\begin{equation}
X_{i,s}^k = 
\begin{cases} 
1 & \text{if the path targeting } k \text{ is located at node } i \text{ at time step } s, \\
0 & \text{if the path targeting } k \text{ is not located at node } i \text{ at time step } s.
\end{cases}
\label{Xijk}
\end{equation}

Specifically, \(X_{i,s}^k = 1\) indicates that the path associated with target \(C_k\) arrives at node \(i\) at time step \(s\), while \(X_{i,s}^k = 0\) indicates that this path does not reach node \(i\) at time step \(s\).

Meanwhile, \(w_{ij}\) denotes an element of the graph's weight matrix, which is constructed per the following rules:

\begin{equation}
w_{ij} = 
\begin{cases} 
0 & \text{if } i = j \text{ and } i \in V, \\
w_{ij} & \text{if an edge exists between nodes } (i, j), \\
\infty & \text{if no edge exists between nodes } (i, j) \ (\infty \text{ acts as a penalty value}).
\end{cases}
\label{wij}
\end{equation}

Constraints for the Steiner tree are specified as follows:

\noindent \textbf{Constraint 1}: This constraint defines the root node of the Steiner tree (herein designated as node 0). It is formulated as:

\begin{equation}
X_{0,0}^{k_1} = 1
\label{h1}
\end{equation}

In  Eq.\ref{h1}, $k_1$denotes the first target point.

This constraint is converted into the penalty function \(H_1\):
\begin{equation}
H_1 = \left(1 - X_{0,1}^{k_1}\right)^2
\label{H_1}
\end{equation}

\noindent \textbf{Constraint 2}: This constraint mandates that the leaf nodes of the Steiner tree must be target nodes. It is expressed as:
\begin{equation}
\forall k,\ X_{k, S}^k = 1
\label{h2}
\end{equation}

In Eq.\ref{h2}, $S$ denotes the maximum time step. Typically, $S = n$, where $n$ represents the size of the vertex set $V$ in graph $G$.

The corresponding penalty function \(H_2\) is:
\begin{equation}
H_2 = \sum_{k} \left(1 - X_{k, S}^k\right)^2
\label{H_2}
\end{equation}

\noindent \textbf{Constraint 3}: This constraint requires that each path can occupy at most one node at any time step. It is given by:
\begin{equation}
\forall k, s,\ \sum_{i} X_{i,s}^k \leq 1
\label{h3}
\end{equation}
The associated penalty function \(H_3\) is:
\begin{equation}
H_3 = \sum_{k, s} \left[ \left(\sum_{i} X_{i,s}^k\right)^2 - \left(\sum_{i} X_{i,s}^k\right) \right]
\label{H_3}
\end{equation}

\noindent \textbf{Constraint 4}: This constraint enforces path continuity: subsequent path segments cannot be empty if earlier segments are occupied. It is formulated as:
\begin{equation}
\forall k, s,\ \sum_{i} X_{i,s}^k \leq \sum_{i} X_{i,s+1}^k
\label{h4}
\end{equation}

The corresponding penalty function \(H_4\) is:

\begin{equation}
H_4 = \sum_{k, s} \left[ \sum_{i} X_{i,s}^k - \left(\sum_{i} X_{i,s}^k\right)\left(\sum_{i} X_{i,s+1}^k\right) \right]
\label{H_4}
\end{equation}

\noindent \textbf{Constraint 5}: This constraint prohibits consecutive stays at non-target nodes. Let \(\mathcal{K}\) denote the set of target nodes; the constraint is expressed as:
\begin{equation}
\sum_{k} \sum_{i \notin \mathcal{K}} \sum_{s} X_{i,s}^k \cdot X_{i,s+1}^k = 0
\label{h5}
\end{equation}

The associated penalty function \(H_5\) is:

\begin{equation}
H_5 = \sum_{k} \sum_{i \notin \mathcal{K}} \sum_{s} X_{i,s}^k \cdot X_{i,s+1}^k
\label{H_5}
\end{equation}

\noindent \textbf{Constraint 6}: This constraint identifies Steiner points: for each target \(k\), its starting node must coincide with a node in the path of another target \(k'\). It is given by:
\begin{equation}
\sum_{i} \sum_{s} \sum_{k \neq k'} X_{i,s}^k \cdot X_{i,s}^{k'} \geq 1
\label{H6}
\end{equation}

The corresponding penalty function \(H_6\) (using a slack variable \(y\) for inequality constraint implementation) is:

\begin{equation}
H_6 = \sum_{k} \left( \sum_{i} \sum_{s} X_{i,s}^k \cdot X_{i,s}^{k'} - y - 1 \right)^2
\label{H_6}
\end{equation}

Where \(y\) denotes a slack variable introduced to convert inequality constraints into quadratic penalty terms, ensuring compatibility with the QUBO framework.

The total penalty function, integrating all constraint violations, is:
\begin{equation}
H_B = H_1 + H_2 + H_3 + H_4 + H_5 + H_6
\label{H_B}
\end{equation}

Where \(H_1\)–\(H_6\) correspond to the penalty functions derived from the six constraints above. Combining the objective function \(H_A = \sum_{i,j,k,s} w_{ij}^k X_{i,s}^k X_{j,s+1}^k\) with the constraint penalty term yields the complete QUBO model:
\begin{equation}
H = H_A + \lambda \cdot H_B
\label{H}
\end{equation}
Here, \(\lambda\) represents a positive penalty coefficient, selected as a sufficiently large value relative to the maximum edge weight in the graph to ensure constraint satisfaction takes priority over objective function minimization.

\maketitle

\section{Experiment}

We utilized the PYQUBO SDK for the simulation and testing of the proposed algorithm. Specifically, we employed \texttt{model.to\_qubo} to convert the algorithm into a Binary Quadratic Model (BQM). This transformation converts constraints into penalty terms, yielding an unconstrained model that can be further processed as the Quadratic Unconstrained Binary Optimization (QUBO) formulation for quantum optimization. It is worth noting that this transformation alters the problem structure, and due to the introduction of bias terms and penalty terms, solutions obtained from the Integer Linear Programming (ILP) and QUBO formulations will have different energy values.

Simulated Quantum Annealing (SQA) was implemented using \texttt{oj.SQASampler} and \texttt{sampler.sample\_qubo} from the PYQUBO library. During execution, the \texttt{num\_reads} parameter of the sampler was set to 1000, while all other parameters were retained at their default values as specified in the official documentation.

\begin{figure}[htbp] 
    \centering 
    \includegraphics[width=0.7\textwidth]{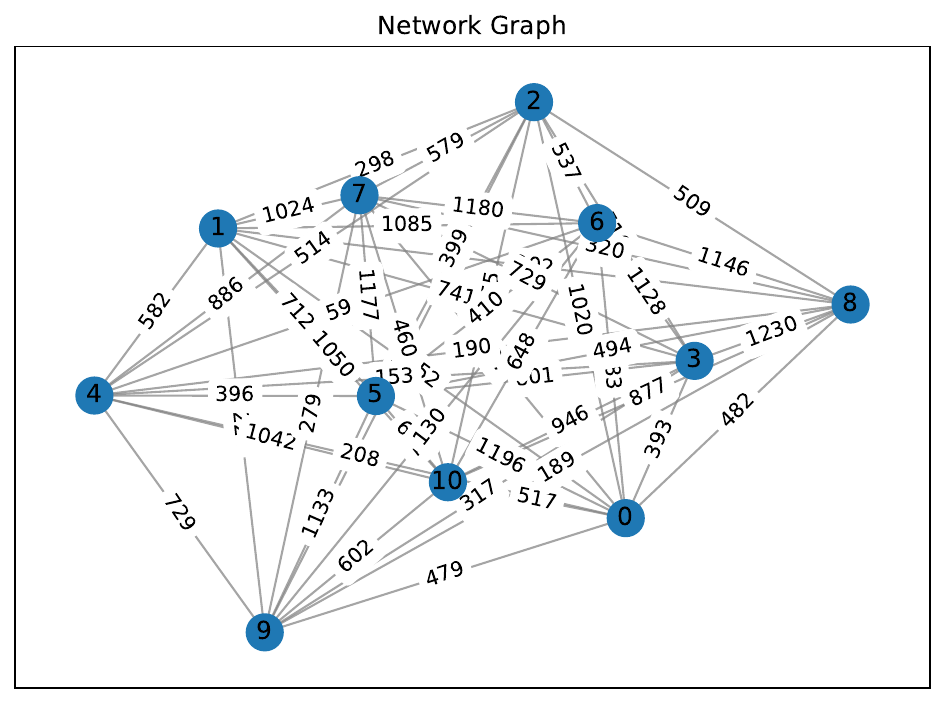} 
    \caption{“Network Graph for Steiner Tree Problem”} 
    \label{fig:network} 
\end{figure}

\begin{figure}[htbp] 
    \centering 
    \includegraphics[width=0.7\textwidth]{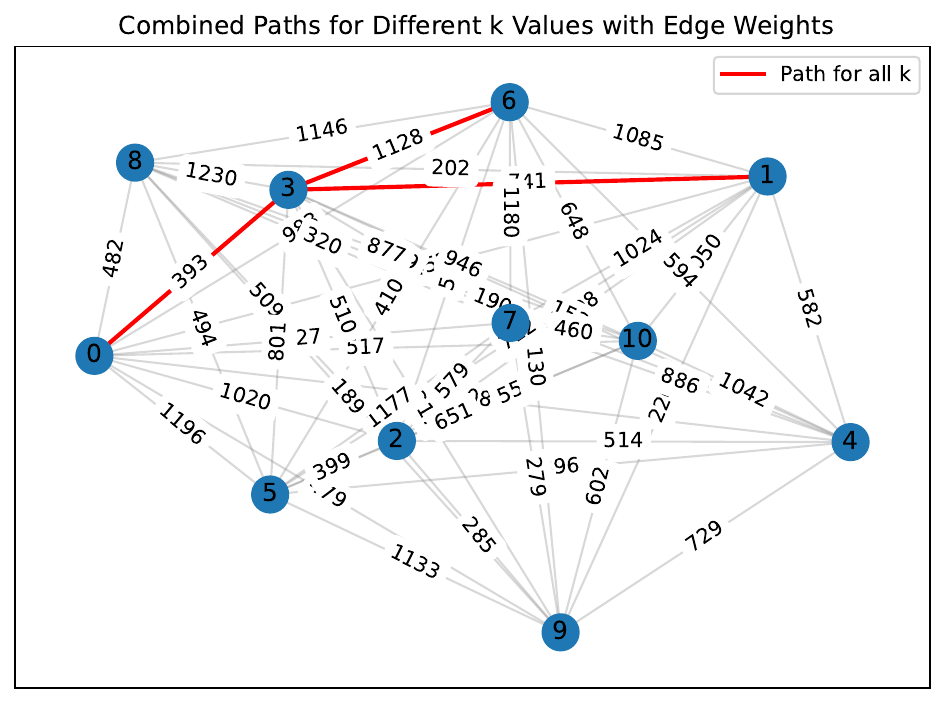} 
    \caption{“Network Graph for Steiner Tree Problem”} 
    \label{fig:path} 
\end{figure}

Experimental tests were conducted on a graph consisting of 11 nodes, as illustrated in Figure \ref{fig:network}. Random weights ranging from 100 to 1000 were assigned to the edges between nodes, and a weight of 10000 was set for non-existent edges. Similarly, the penalty value for constraint violations was set to 10000. After 10 independent runs, the collected data were processed, and the corresponding results are presented in Figure \ref{fig:path}.

Through analysis, we found that when the number of target nodes is small, the proposed algorithm can efficiently find the Minimum Steiner Tree (MST) via SQA. 

\maketitle

\section{Conclusions }

The minimum Steiner tree problem is a fundamental challenge in combinatorial optimization, with critical applications in network topology design, integrated circuit layout, and distributed resource allocation. While classical approaches—including exact, approximation, and heuristic algorithms—are widely used, they suffer from exponential time complexity as the graph size increases, severely limiting their scalability for large-scale instances.

To tackle this limitation, we utilize quantum annealing techniques: we formulate the minimum Steiner tree problem into a Quadratic Unconstrained Binary Optimization (QUBO) model by encoding core constraints (e.g., node connectivity, path continuity, and Steiner point identification) into penalty functions, enabling the problem to be deeply compatible with quantum computing’s solution-space exploration capabilities.

This approach not only aims to find optimal Steiner tree solutions for combinatorial optimization tasks but also significantly reduces the computational overhead of traditional classical methods. Our relevant analysis results demonstrate that the approach relies centrally on the unique computational properties of quantum systems (e.g., quantum superposition and quantum tunneling) to enhance the efficiency of combinatorial optimization, thereby effectively finding high-quality solutions for the minimum Steiner tree problem—especially in large-scale instances that are intractable for traditional classical algorithms, where the advantages of quantum computing are even more pronounced.

\bibliographystyle{unsrt} 
\bibliography{references} 

\end{document}